\documentclass[12pt]{article}

\usepackage{latexsym}
\usepackage{amsmath}
\usepackage{multibox}
\usepackage{verbatim}
\usepackage{mathrsfs}

\usepackage{amssymb,amsmath}
\usepackage{amsthm}
\usepackage{amscd}
\usepackage{amssymb}
\usepackage{amsfonts}
\usepackage{url}
\usepackage{slashed}
\usepackage[latin1]{inputenc}
\usepackage[english]{babel}
\usepackage{putex}

 \csname
@addtoreset\endcsname{equation}{section}

\def\gz0{\gamma^{0}}

\def\scs#1{\section{\sc #1}}
\def\scss#1{\subsection{\sc #1}}

\def\m{\mu}
\def\n{\nu}

\def\p{\pi}

\newcommand{\beq}{\begin{equation}}
\newcommand{\eeq}[1]{\label{#1}\end{equation}}
\newcommand{\bea}{\begin{eqnarray}}
\newcommand{\eea}[1]{\label{#1}\end{eqnarray}}

\def\bs{\begin{split}}
\def\es{\end{split}}
\def\ba{\begin{array}}
\def\ea{\end{array}}
\def\bec{\begin{center}}
\def\ec{\end{center}}
\def\ba{\begin{align}}
\def\ena{\end{align}}

\def\12{\frac{1}{2}}

\begin{document}

\begin{flushright}
{\today}
\end{flushright}

\vspace{25pt}

\begin{center}
%%%%%%%%%%%%%%%%%%%%%%%%%%%%%%%%%%%%%%%%%%%%%%%%%%%%%%%%%%%%%%%%%%%%
{\Large\sc Notes on a Cure for Higher-Spin Acausality}\\
%%%%%%%%%%%%%%%%%%%%%%%%%%%%%%%%%%%%%%%%%%%%%%%%%%%%%%%%%%%%%%%%%%%%
\vspace{25pt}
{\sc Massimo~Porrati$^a$ and Rakibur~Rahman$^b$}\\[15pt]
{\sl\small
$a$) Center for Cosmology and Particle Physics\\
    Department of Physics, New York University\\
    4 Washington Place, New York, NY 10003, USA\\
\vspace{6pt}
$b$) Scuola Normale Superiore and INFN\\
Piazza dei Cavalieri, 7\\I-56126 Pisa \ ITALY \\
\vspace{6pt}
e-mail: {\small \it
massimo.porrati@nyu.edu, rakibur.rahman@sns.it}}\vspace{10pt}
%%%%%%%%%%%%%%%%%%%%%%%%%%%%%%%%%%%%%%%%%%%%%%%
\vspace{35pt}

{\sc\large Abstract}

\end{center}
%%%%%%%%%%%%%%%%%%%%%%%%%%%%%%%%%%%%%%%%%%%%%%%
We present a Lagrangian describing a massive charged spin-2 field and a scalar in a constant electromagnetic background,
and we provide a consistent description of the system. The Lagrangian, derived from string field theory through a suitable
dimensional reduction, propagates the correct number of degrees of freedom within the light cone in any space-time dimension
less than 26. We briefly discuss the higher-spin generalization of this construction, that cures the pathologies of a massive
charged particle of arbitrary integer spin by introducing only finitely many new massive degrees of freedom.

\setcounter{page}{1}
\pagebreak

%%%%%%%%%%%%%%%%%%%%%%%%%%%%%%%%%%%
\scs{Introduction}\label{sec:intro}
%%%%%%%%%%%%%%%%%%%%%%%%%%%%%%%%%%%

Electromagnetic (EM) or gravitational interactions generally produce severe inconsistencies in local actions that
attempt to describe massive fields with spin larger than one, not only when EM and gravity are dynamical, but even
when these interactions are treated as external classical backgrounds~\cite{VZ,Others,Deser}. The interacting
Lagrangian may not propagate the correct number of degrees of freedom (DoF), or may allow the physical modes to move
faster than light.  Such pathologies indeed originate from the presence of high-spin gauge modes~\cite{PR1}.

Even the very simple interaction setup of a constant external EM background is fraught with difficulties.
A well-known example is that of a massive charged spin-2 field, for which, although there exists a unique minimally
coupled Lagrangian that preserves the DoF count~\cite{F}, the resulting modes suffer from a ``Velo-Zwanziger
acausality"~\cite{VZ}, i.e. they cease to propagate within the light cone. This problem persists for a wide class
of non-minimal models~\cite{Deser} making it quite challenging, field theoretically, to construct consistent EM
interactions for a charged massive spin-2 particle (and generically for any higher-spin field).

However, it is not impossible to restore causal propagation. In fact, there are a number of explicit examples
where such difficulties are circumvented either through the addition of new dynamical DoFs or through appropriate
non-minimal terms. For instance, the $\mathcal N = 2$ gauged supergravity~\cite{fpvn}, with supersymmetry broken
without a cosmological constant~\cite{broken}, contains a massive spin-3/2 gravitino that can be charged under
the graviphoton and still propagate consistently. Here, it is the presence of dynamical gravity
that rescues causality for the charged gravitino~\cite{Deser}. On the other hand, ref.~\cite{PR3} presented a
consistent non-minimal model containing a \emph{single} massive spin-3/2 field propagating in a constant EM background.

String theory also bypasses the Velo-Zwanziger problem (at least in a constant EM background)
for an arbitrary integer spin $s$, in that any field
belonging to the first Regge trajectory of the open bosonic string interacts consistently and causally with a
constant EM background~\cite{AN,PRS}, thanks to the highly non-minimal (kinetic) terms that the theory itself
spells out. While the explicit string-theoretic Lagrangian for spin 2 was given in~\cite{AN}, and for any higher
spin the corresponding non-minimal Lagrangian is guaranteed to exist by the results of ref.~\cite{PRS}, these Lagrangians do not hold
good in arbitrary space-time dimensions. Unlike the free-string case, a unitary-gauge Lagrangian in
the presence of the background is consistent only in the critical dimension $D = 26$. The same is
true for the spin-2 Argyres-Nappi Lagrangian~\cite{AN}, as was shown in~\cite{PRS}.

Of course, one can perform a dimensional reduction of the 26-dimensional Lagrangian provided by String Theory for
an arbitrary integer spin $s$, to a space-time dimension $d<26$. The $d$-dimensional model obtained
thereby would be a system of all integer spins from $s$ down to 0, coupled to a constant EM background, with
each field propagating causally. The spin-$s$ acausality in $d$ dimensions would then be cured by the presence
of additional dynamical fields with spins $s-1$, $s-2$, ..., 0.
The main point presented in this paper, however, is that one can consistently throw away the non-singlets of
the $SO(26-d)$ symmetry rotating the internal coordinates. By this reduction, one ends up with a $d$-dimensional
model that contains just the spins $s, s-2, s-4, ..., \tfrac{1}{2}\left[1+(-)^{s+1}\right]$, all of which propagate
within the light cone with the correct DoFs. In particular, such a reduction of the Argyres-Nappi Lagrangian~\cite{AN}
gives rise to a coupled system of a spin 2 and a scalar that is consistent in any space-time dimension less than 26.
Thus one \emph{can} restore causality for a massive charged spin 2, in $d = 4$ for example, just by adding a dynamical
scalar along with non-minimal terms.

The organization of this paper is as follows. In Section~\ref{sec:spin2} we construct the aforementioned model
for a massive charged spin-2 field and show its consistency: in particular, Section~\ref{sec:AN} presents the
Argyres-Nappi spin-2 Lagrangian with some necessary details, while Section~\ref{sec:Reduction} carries out the
dimensional reduction for obtaining the actual model involving a spin 2 and a scalar. Section~\ref{sec:Cons}
proves the consistency of the model for arbitrary dimensions less than 26, whereas Section~\ref{sec:noghost}
proves a crucial property of the dimensional reduction procedure, i.e. that it is free of negative-norm states.
We discuss similar constructions for higher integer spins in Section~\ref{sec:arb}, focusing on the equations of
motion (EoM) for simplicity. Finally, we make some concluding remarks in Section~\ref{sec:Conclusion}.

\vskip 36pt

%%%%%%%%%%%%%%%%%%%%%%%%%%%%%%%%%%%%%%%%%%%%%%%%%%%%%%%%%%%%%%%%%%%%
\scs{A Consistent Model for Charged Massive Spin 2}\label{sec:spin2}
%%%%%%%%%%%%%%%%%%%%%%%%%%%%%%%%%%%%%%%%%%%%%%%%%%%%%%%%%%%%%%%%%%%%

In this Section, we dimensionally reduce the spin-2 Argyres-Nappi Lagrangian~\cite{AN} to construct
a model containing a spin 2 and a scalar, and show that it indeed propagates both fields causally
with the correct number of DoFs in any $d<26$.

\vskip 24pt

%%%%%%%%%%%%%%%%%%%%%%%%%%%%%%%%%%%%%%%%%%%%%%%%%%%%%%%%
\scss{The Spin-2 Argyres-Nappi Lagrangian}\label{sec:AN}
%%%%%%%%%%%%%%%%%%%%%%%%%%%%%%%%%%%%%%%%%%%%%%%%%%%%%%%%

The Argyres-Nappi Lagrangian~\cite{AN}, which describes a massive charged spin-2 field in a constant
EM background, is given by
\bea L_{\text{AN}}&=&\mathcal{H}_{\mu\nu}^*\left(\mathfrak{D}^2
-2-\tfrac{1}{2}\text{Tr}G^2\right) h^{\mu\nu}-2i\mathcal{H}_{\mu\nu}^*(G\cdot h-h\cdot G)
^{\mu\nu}-\mathcal{H}^*\left(\mathfrak{D}^2-2-\tfrac{1}{2}\text{Tr}G^2\right)\mathcal{H}\nonumber\\&&-\,\mathcal{H}_
{\mu\nu}^*\left\{\mathfrak D^\mu\mathfrak D^\rho[\,(1+iG)\cdot h\,]_\rho^{~\nu}-\tfrac{1}{2}\,\mathfrak D^\mu\mathfrak
D^\nu\mathcal H+(\mu\leftrightarrow\nu)\right\}+\mathcal{H}^*\mathfrak D^\mu\mathfrak D^\nu\mathcal H_{\mu\nu},
\eea{AN}
where the mass-squared\footnote{Here we are talking about the mass that appears in the free theory, i.e.~in the
absence of the EM background. As it is well known, the constant background causes a shift in the mass~\cite{AN,PRS}.}
of the spin-2 field has been set equal to 2,  and $\mathfrak D^\mu$ is related to the covariant
derivative $\mathcal D^\mu$ as \beq \mathfrak D^\mu=\left(\sqrt{G/eF}\right)^{\m\n}\mathcal D_\n,\qquad
[\,\mathfrak D^\m, \mathfrak D^\n\,]=-iG^{\m\n},\qquad [\,\mathcal D^\m, \mathcal D^\n\,]=-ieF^{\m\n}.\eeq{covder}
The charge of the spin-2 field is $e$, and $F^{\m\n}$ is the EM field strength. Moreover,
\beq \mathcal H_{\mu\nu}\equiv(1+iG)_\mu^{~\alpha}
(1+iG)_\nu^{~\beta}\,h_{\alpha\beta},\qquad \mathcal{H}\equiv\mathcal{H}^\mu_{~\mu}.
\eeq{Hh}
The antisymmetric rank-2 Lorentz tensor $G^{\m\n}$ is given by the expression
\beq
G=\frac{1}{\pi}\left[\,\tanh^{-1}(\pi e_0F)+\tanh^{-1}(\pi e_\p F)\,\right],
\eeq{G}
which reflects the fact that the Lagrangian~(\ref{AN}) has been derived from the theory of open bosonic strings, as
it contains in it the string endpoint charges $e_0$ and $e_\pi$, with $e\equiv e_0+e_\pi$ being the total string
charge.

The Lagrangian~(\ref{AN}) is Hermitian, and gives rise to the spin-2 Fierz-Pauli system~\cite{AN,PRS}
\beq \left(\mathfrak{D}^2-2-\tfrac{1}{2}\text{Tr}G^2\right)\mathcal{H}_{\mu\nu}-2i(G\cdot\mathcal{H}-\mathcal{H}\cdot G)
_{\mu\nu}=0,\qquad \mathfrak D^\mu\mathcal H_{\mu\nu}=0,\qquad\mathcal H=0,
\eeq{FPsystem}
which preserves the right number of DoFs, namely $\12(D+1)(D-2)$ in $D$ space-time dimensions, and admits only
causal propagation~\cite{AN}.

However, as already mentioned in the Introduction, the Fierz-Pauli system~(\ref{FPsystem}) follows from the
Argyres-Nappi Lagrangian~(\ref{AN}) \emph{only} when $D = 26$~\cite{AN,PRS}. Away from the critical dimension,
the trace of $\mathcal H_{\mu\nu}$  becomes dynamical~\cite{PRS}, so that one has more propagating DoFs than those
of a massive spin-2 field. Therefore, the Lagrangian~(\ref{AN}) does \emph{not} cure the original Velo-Zwanziger
problem for arbitrary space-time dimensions.

As was first noted in~\cite{AN}, the consistency of the Argyres-Nappi Lagrangian~(\ref{AN}) is not affected by the
substitution:
\beq
G^{\m\n}\rightarrow eF^{\m\n},\qquad \mathfrak D^\mu \rightarrow \mathcal D^\mu.
\eeq{subs}
In what follows we will use interchangeably $\mathfrak D^\mu$ with $\mathcal D^\mu$, and $G^{\m\n}$ with
$eF^{\m\n}$. Thus, all our subsequent expressions will contain a single charge $e$, as it is suitable for a point particle.

\vskip 24pt

%%%%%%%%%%%%%%%%%%%%%%%%%%%%%%%%%%%%%%%%%%%%%%%%%%%%%%%%%%%%%%%%%%%%%%%%%%%%%%%%%
\scss{Dimensional Reduction of The Argyres-Nappi Lagrangian}\label{sec:Reduction}
%%%%%%%%%%%%%%%%%%%%%%%%%%%%%%%%%%%%%%%%%%%%%%%%%%%%%%%%%%%%%%%%%%%%%%%%%%%%%%%%%

We would like to reduce the Argyres-Nappi Lagrangian~(\ref{AN}) consistently from $D=26$ space-time dimensions
to a $d$-dimensional model, where, obviously, $d<D$. With this end in view, we split the $D$ space-time coordinates,
labeled by Greek indices, into two subsets:
\bea
\m=\left\{
            \begin{array}{ll}
              m=0,\,1,\,2,\,...\,,\,d-1,\\
              M=d,\,d+1,\, ...\,,\,D-1.
            \end{array}
          \right.
\eea{label}
That is, the $d$-dimensional space-time coordinates are labeled by lower-case Roman indices, while the remaining
($D-d$) spatial ones, which we will consider as internal coordinates, are labeled by upper-case Roman indices.

The dimensional reduction consists in keeping only those fields that are singlets of the internal coordinates.
Thus, the $D$-dimensional fields $h_{\m\n}$ and $G_{\m\n}$  split as
\beq
h_{\m\n}=\left(
\begin{aligned}
\begin{tabular}{c|c}
  $h_{mn}$ & $0$\\
  \hline
  $0$ & $\tfrac{1}{D-d}\,\delta_{MN}\,\phi$\\
\end{tabular}
\end{aligned}
\right),\qquad
G_{\m\n}=\left(
\begin{aligned}
\begin{tabular}{c|c}
  $G_{mn}$ & $0$\\
  \hline
  $0$ & $0$\\
\end{tabular}
\end{aligned}
\right),
\eeq{hGsplit}
so that, in view of the first equation in~(\ref{Hh}), $\mathcal H_{\m\n}$ has the reduction
\beq
\mathcal H_{\m\n}=\left(
\begin{aligned}
\begin{tabular}{c|c}
  $[1+iG]_{ma}[1+iG]_{nb}\,h^{ab}$ & $0$\\
  \hline
  $0$ & $\tfrac{1}{D-d}\,\delta_{MN}\,\phi$\\
\end{tabular}
\end{aligned}
\right) \equiv \left(
\begin{aligned}
\begin{tabular}{c|c}
  $\mathfrak{h}_{mn}$ & $0$\\
  \hline
  $0$ & $\tfrac{1}{D-d}\delta_{MN}\phi$\\
\end{tabular}
\end{aligned}
\right).
\eeq{Hsplit}
All the fields obtained thereby are assumed to be functions of the $d$ space-time coordinates only, and not
of the internal coordinates; moreover, the EM background is nonzero only in $d$ dimensions. This means
that the covariant derivative splits as
\beq
\mathcal D^{\m}=\left(
\begin{aligned}
\begin{tabular}{cc}
  $\mathcal D^m$ \\
  \hline
  $0$\\
\end{tabular}
\end{aligned}
\right).
\eeq{Dsplit}
The Lagrangian~(\ref{AN}) then reduces to
\bea
L_d&=&\mathfrak{h}_{mn}^*\left(\mathcal{D}^2-2-\tfrac{1}{2}\text{Tr}G^2\right) h^{mn}-2i\mathfrak{h}_{mn}^*
(G\cdot h-h\cdot G)^{mn}-\mathfrak{h}^*\left(\mathcal{D}^2-2-\tfrac{1}{2}\text{Tr}G^2\right)\mathfrak{h}\nonumber\\&&
-\mathfrak{h}_{mn}^*\left\{\mathcal D^m\mathcal D^p[\,(1+iG)\cdot h\,]_p^{~n}-\tfrac{1}{2}\,\mathcal D^m\mathcal
D^n\mathfrak h+(m\leftrightarrow n)\right\}+\mathfrak{h}^*\mathcal D^m\mathcal D^n\mathfrak h_{mn}\nonumber\\&&
+\left[\,\mathfrak{h}_{mn}^*\mathcal D^m\mathcal D^n\phi-\left\{\mathfrak{h}^*+\tfrac{1}{2}\left(\tfrac{D-d-1}{D-d}
\right)\phi^*\right\}\left(\mathcal{D}^2-2-\tfrac{1}{2}\text{Tr}G^2\right)\phi\,+\,\text{h.c.}\,\right],\eea{model}
where $d$ is the space-time dimensionality, $D = 26$ by construction, and $\mathfrak h^m_{~m}$ has been denoted
by $\mathfrak h$.

Eq.~(\ref{model}) is our desired Lagrangian. We claim that, up to local field redefinitions, it consistently describes
a charged massive spin-2 field coupled to a charged massive scalar in a constant external EM background in any space-time
dimension $d<26$.

It is not difficult to see that the Lagrangian~(\ref{model}) gives a decoupled system of a spin 2 plus
a scalar in the absence of a background field. Indeed, the field redefinition
\beq
\mathfrak h_{mn} \rightarrow \mathfrak h_{mn}-\left(\tfrac{1}{d-1}\right)\left[\eta_{mn}\phi-\tfrac{1}{4}
\mathcal D_{(m}\mathcal D_{n)}\phi\,\right]
\eeq{redef}
reduces the Lagrangian to
\bea
L_d&\rightarrow&\mathfrak{h}_{mn}^*\left(\mathcal{D}^2-2\right)\mathfrak{h}^{mn}-\mathfrak{h}^*\left(
\mathcal{D}^2-2\right)\mathfrak{h}-\mathfrak{h}_{mn}^*\mathcal D^{(m}\left[\mathcal D\cdot\mathfrak h\right]^{n)}
+\mathfrak{h}_{mn}^*\mathcal D^m\mathcal D^n\mathfrak h+\mathfrak{h}^*\mathcal D^m\mathcal D^n\mathfrak h_{mn}\nonumber\\
&&+\left[\tfrac{D-1}{(d-1)(D-d)}\right]\phi^*\left(\mathcal{D}^2-2\right)\phi\,+\,\mathcal{O}(G),
\eea{redefLag}
so that in the limit $G \rightarrow 0$, one obtains the Fierz-Pauli Lagrangian for the massive spin-2 field
$\mathfrak h_{mn}$, decoupled from the massive scalar $\phi$. At $\mathcal O(G)$ there appear (kinetic)
mixings of the two fields, and these are precisely what make the model consistent\footnote{We postpone for later
the possibility that $\mathcal O(G)$ corrections to the redefinition~(\ref{redef}) may eliminate these mixings,
so as to give a description where either of the fields propagates consistently in isolation.}, as we will see below.

\vskip 24pt

%%%%%%%%%%%%%%%%%%%%%%%%%%%%%%%%%%%%%%%%%%%
\scss{Proof of Consistency}\label{sec:Cons}
%%%%%%%%%%%%%%%%%%%%%%%%%%%%%%%%%%%%%%%%%%%

The variation of the Lagrangian~(\ref{model}) gives rise to the following EoMs.
\bea
\mathcal R_{mn}&\equiv&\left(\mathcal
{D}^2-2-\tfrac{1}{2}\text{Tr}G^2\right)\left[\,\mathfrak{h}_{mn}-(1+G^2)_{mn}(\mathfrak{h}+\phi)\,\right]
-2i(G\cdot\mathfrak{h}-\mathfrak{h}\cdot G)_{mn}\nonumber\\&&+\tfrac{1}{2}\left\{[(1+iG)
\cdot\mathcal D ]_m [ (1+iG)\cdot\mathcal D ]_n+[(1+iG)\cdot\mathcal D]_n [(1+iG)\cdot\mathcal D]_m\right\}
(\mathfrak{h}+\phi)\nonumber\\
&&-\left\{[\,(1+iG)\cdot\mathcal D\,]_m\,\mathcal D^p\mathfrak{h}_{pn}+[\,(1+iG)\cdot
\mathcal D\,]_n\,\mathcal D^p\mathfrak{h}_{pm}\right\}+(1+G^2)_{mn}\mathcal D^a\mathcal D^b
\mathfrak h_{ab}\nonumber\\&=&0,
\eea{a2}
and
\beq
r \equiv \left(\mathcal D^a\mathcal D^b\mathfrak h_{ab}
-\mathcal{D}^2\mathfrak{h}\right)+\left(2+\tfrac{1}{2}\text{Tr}G^2\right)\mathfrak{h}-\left(\tfrac{D-d-1}{D-d}\right)
\left(\mathcal{D}^2-2-\tfrac{1}{2}\text{Tr}G^2\right)\phi=0.
\eeq{a3}
To derive the necessary constraints from the system of equations~(\ref{a2})--(\ref{a3}) one needs to perform some
algebraic manipulations, similar to those presented in Section 6.1 of~\cite{PRS}. One can compute the quantity
$\mathcal R^m_{~m}+\left[2\mathcal D\cdot(1+iG)^{-1}\right]^n\mathcal D^m\mathcal R_{mn}$, from Eq.~(\ref{a2}),
to find
\bea
&(d-6+2\text{Tr}G^2)(\mathcal D^a\mathcal D^b\mathfrak h_{ab}-\mathcal D^2\mathfrak h)+\left[2(d-1)
+\tfrac{1}{2}\text{Tr}G^2(d+4+2\text{Tr}G^2)\,\right]\mathfrak h&\nonumber\\
&-(d-5+2\text{Tr}G^2)\,\mathcal{D}^2\phi
+\left[\,2d+\tfrac{1}{2}(d+5)\text{Tr}G^2+(\text{Tr}G^2)^2\right]\phi=0.&
\eea{a7}
Then one can eliminate $(\mathcal D^a\mathcal D^b\mathfrak h_{ab}-\mathcal D^2\mathfrak h)$ from the above formula
by using Eq.~(\ref{a3}). The result is
\beq \mathfrak h=-\phi+\left[\tfrac{D-6+2\text{Tr}G^2}{(D-d)(10+\text{Tr}G^2)}\right]\left(\mathcal{D}^2-2
-\tfrac{1}{2}\text{Tr}G^2\right)\phi.\eeq{a99}
Also, the quantity $\left[\mathcal D\cdot\{(1+iG)(2+iG)\}^{-1}\right]^n\mathcal D^m\mathcal R_{mn}$ gives from
Eq.~(\ref{a2})
\bea &-\left[1+\left(\tfrac{iG}{2+iG}\right)^{mn}\mathcal D_m\mathcal D_n\right](\mathcal D^a\mathcal D^b
\mathfrak h_{ab}-\mathcal D^2\mathfrak h)-\left[\tfrac{1}{4}\text{Tr}G^2+\tfrac{1}{2}(5+\text{Tr}G^2)\left(\tfrac
{iG}{2+iG}\right)^{mn}\mathcal D_m\mathcal D_n\right]\mathfrak h&\nonumber\\&+\left(\tfrac{iG}{2+iG}\right)
^{mn}\mathcal D_m\mathcal D_n\left(\mathcal{D}^2-\tfrac{5}{2}-\tfrac{1}{2}\text{Tr}G^2\right)\phi+\left(
\mathcal{D}^2-\tfrac{1}{4}\text{Tr}G^2\right)\phi~=~0.&\eea{a8} In the above, Eq.~(\ref{a3}) can be used once
again to eliminate $(\mathcal D^\alpha\mathcal D^\beta\mathfrak h_{\alpha\beta}-\mathcal D^2\mathfrak h)$, and
then Eq.~(\ref{a99}) removes $\mathfrak h$ from the resulting expression, giving thereby
\bea 0&=&\left[\,2(D-1)+\tfrac{1}{4}\,\text{Tr}G^2(D+14+2\,\text{Tr}G^2)\,\right]\left(\mathcal{D}^2-2-\tfrac{1}{2}
\text{Tr}G^2\right)\phi\nonumber\\&&-\tfrac{1}{2}(D-26)\left(\tfrac{iG}{2+iG}\right)^{mn}\mathcal D_m\mathcal D_n
\left(\mathcal{D}^2-2-\tfrac{1}{2}\text{Tr}G^2\right)\phi.\eea{a9}
Because $D = 26$ by construction, this gives a second order equation for $\phi$, namely
\beq \left(\mathcal{D}^2-2-\tfrac{1}{2}\text{Tr}G^2\right)\phi=0,\eeq{a100} which, in turn, implies from Eq.~(\ref{a99})
that \beq \mathfrak h=-\phi.\eeq{a101} Eqs.~(\ref{a100}) and~(\ref{a101}), when used in the EoM~(\ref{a3}), requires that
$\mathcal D^a\mathcal D^b\mathfrak h_{ab}$ vanishes. The divergence equation, $\mathcal D^m\mathcal R_{mn}=0$, i.e.
\bea 0&=&-\left\{(1+iG)(2+iG)\right\}_n^{~a}\mathcal D^b\mathfrak h_{ab}-\left\{iG(1+iG)\right\}
_n^{~s}\mathcal D_s(\mathcal D^a\mathcal D^b \mathfrak h_{ab})\nonumber\\&&+\left\{(1+iG)
\left[\left(2-\tfrac{1}{2}iG+\tfrac{3}{2}G^2\right)+iG\left(\mathcal D^2-\tfrac{1}{2}\text{Tr}G^2\right)\right]\right\}
_n^{~a}\mathcal D_a(\mathfrak h+\phi),\nonumber\eea{a1000} now obviously implies the divergence constraint
\beq \mathcal D^m\mathfrak h_{mn}=0,\eeq{a103} thanks to Eq.~(\ref{a101}) and the fact that
$\mathcal D^a\mathcal D^b\mathfrak h_{ab}$ vanishes. Finally, in view of Eqs.~(\ref{a101}) and~(\ref{a103}), one
finds from~(\ref{a2}) that \beq \left(\mathcal{D}^2-2-\tfrac{1}{2}\text{Tr}G^2\right)\mathfrak h_{mn}
-2i(G\cdot\mathfrak h-\mathfrak h\cdot G)_{mn}=0.\eeq{a104}
Eqs.~(\ref{a100})--(\ref{a104}) are an algebraically consistent set of equations. The divergence and trace constraints
ensure the correct DoF count for the spin-2 field $\mathfrak h_{mn}$, whose propagation, as well as that of $\phi$,
is causal, as is manifest from Eqs.~(\ref{a100}) and~(\ref{a104}). Thus Eqs.~(\ref{a100})--(\ref{a104})
consistently describe a massive charged spin-2 field plus a massive charged scalar
in a constant EM background, and they follow from the Lagrangian~(\ref{model}).

If one takes the model~(\ref{model}) at face value, without knowing about its string-theoretic origin, one would
discover, from the analysis presented in this Section, that it provides a consistent description of the fields
$\mathfrak h_{mn}$ and $\phi$ only when the parameter $D$ is set equal to 26. Because the scalar kinetic term appearing
in~(\ref{redefLag}) has the wrong sign for $(D-d)<0$, the model is clearly plagued with a ghost, and therefore
does not hold good, when $d>26$. To see what happens at $d = 26$, we can of course notice that~(\ref{model}) comes from
Eq.~(\ref{AN}), which contains no scalar. Alternatively, we can redefine $\phi\rightarrow\sqrt{(D-d)}\,\phi$\,, to make~(\ref{model})
non-singular in the limit $d\rightarrow26$. In this limit, the scalar field, which is now canonically normalized, completely
decouples from the spin-2 Lagrangian.

As one expects, Eqs.~(\ref{a100})--(\ref{a104}) can also be derived directly from dimensional reduction of
the Fierz-Pauli system~(\ref{FPsystem}). We will use this important fact in Section~\ref{sec:arb}, where we will consider
fields with arbitrary integer spin, for which we do not have any explicit Lagrangians at our disposal. Notice that
the implementation of the field redefinition~(\ref{redef}) on the Eqs.~(\ref{a100})--(\ref{a104}), gives, after some
algebraic manipulations, the system \bea &\left(\mathcal{D}^2-2-\tfrac{1}{2}\text{Tr}G^2\right)
\mathfrak h_{mn}-2i(G\cdot\mathfrak h-\mathfrak h\cdot G)_{mn}=0,&\label{new1}\\&\mathcal D^m\mathfrak h_{mn}
=-\tfrac{1}{4}\left(\tfrac{1}{d-1}\right)\left(3iG_{na}+\eta_{na}\text{Tr}G^2\right)\mathcal D^a\phi,&\label{new2}\\
&\mathfrak h=-\tfrac{1}{4}\left(\tfrac{1}{d-1}\right)\text{Tr}G^2\,\phi,&\label{new3}\\&\left(\mathcal{D}^2-2-
\tfrac{1}{2}\text{Tr}G^2\right)\phi=0.&\eea{new4}
This again shows that, when $G\rightarrow0$, the spin-2 field decouples from the scalar.

\vskip 24pt

%%%%%%%%%%%%%%%%%%%%%%%%%%%%%%%%%%%%%%%%%%%%%%%%%%%%%%%%%%%%%%%
\scss{Non-Existence of Negative-Norm States}\label{sec:noghost}
%%%%%%%%%%%%%%%%%%%%%%%%%%%%%%%%%%%%%%%%%%%%%%%%%%%%%%%%%%%%%%%

It was shown in~\cite{PRS} that in a constant EM background the charged open-string spectrum is free of negative-norm
states. One may wonder whether a dimensional reduction of the system that throws away some of the fields (non-singlets
of the internal coordinates) could introduce ghosts. As we will see now, for $SO(26-d)$-singlet states obtained
in the dimensional reduction of the bosonic string to $d$ dimensions, the $D=26$ no-ghost theorem also guarantees the
absence of negative-norm states for $d<26$.

Because the BRST charge $Q$ commutes with $SO(26-d)$, it maps the $SO(26-d)$-singlet sector of the string Hilbert
space $V$ into itself. More generally, if one decomposes $V$ into $SO(26-d)$ irreps as $V=\sum_I V_I$,
then $Q(V_I) \subset V_I$. In other words, $Q$ acts ``diagonally" on the irreps. In particular, let us consider a
$Q$-closed but spurious state $v=Qu$, where we decompose $v=\sum_I v_I,~v_I \in V_I$. The diagonal
action of $Q$ means that $v_I=Qu_I,~u_I \in V_I$. So, in particular, if a spurious $v_0$ is a singlet,
$v_0 \in  V_0$ ($I=0$ labels the singlet subspace), then $v_0$ is the image of a singlet. Likewise, if $v$
is $Q$-closed, then each of the component $v_I$'s are also $Q$-closed: $Qv_I=0$.

Therefore, the $Q$-cohomology decomposes into a direct sum of cohomologies, each of which is a positive-norm Hilbert space,
with a metric that is the restriction of the metric on $V$ to $V_I$ (since the metric is an $SO(26)$ singlet).

\vskip 36pt

%%%%%%%%%%%%%%%%%%%%%%%%%%%%%%%%%%%%%%%%%%%
\scs{Arbitrary Integer Spin}\label{sec:arb}
%%%%%%%%%%%%%%%%%%%%%%%%%%%%%%%%%%%%%%%%%%%

As we already mentioned, String Theory provides a consistent description, at least in 26 space-time dimensions,
of a massive field of arbitrary integer spin $s$, propagating in a constant EM background without the presence of
other dynamical fields~\cite{PRS}. In this case, the generalized Fierz-Pauli conditions~\cite{PRS} read:
\bea &\left[\mathcal{D}^2-2(s-1)-\tfrac{1}{2}\text{Tr}G^2\right]\varPhi_{\mu_1...\mu_s}+2iG^\alpha{}_{~(\mu_1}
\varPhi_{\mu_2...\mu_s)\alpha}=0,&\label{b60}\\&\mathcal{D}^\mu\varPhi_{\mu\mu_2...\mu_s}=0,&\label{b61}\\
&\varPhi^\mu_{~\mu\mu_3...\mu_s}=0,&\eea{b62} in the units in which the mass-squared of the free theory is set
equal to $2(s-1)$.

The singlets of the internal $(26-d)$ coordinates, that survive the dimensional reduction considered here,
are symmetric tensors of rank $s, s-2, s-4, ..., \tfrac{1}{2}\left[1+(-)^{s+1}\right]$. That is, if $s$ is even(odd),
one will end up having all fields with even(odd) spins, from $s$ down to 0(1). Let us denote the field with spin
$(s-2k)$ by $\phi_{s-2k}$, where $0\leq k\leq\tfrac{1}{4}\left[2s-1+(-)^{s}\right]$.
The trace of such a field will be denoted with a prime. The Fierz-Pauli system~(\ref{b60})--(\ref{b62}) then reduces to
\bea &\left[\mathcal{D}^2-2(s-1)-\tfrac{1}{2}\text{Tr}G^2\right]\phi_{s-2k}-2i\left[\,G\cdot\phi_{s-2k}\,\right]_
{\text{symmetrized}}=0,&\label{b63}\\&\mathcal{D}\cdot\phi_{s-2l}=0,\qquad 0\leq l\leq\tfrac{1}{4}\left[2s-3+(-)^{s+1}
\right],&\label{b64}\\&\phi'_{s-2m}=-\phi_{s-2m-2},\qquad 0\leq m\leq\tfrac{1}{4}\left[2s-5+(-)^{s}\right].&\eea{b65}
The system~(\ref{b63})--(\ref{b65}) is, of course, algebraically consistent. The divergence of any field with spin\;$\geq$\;1
vanishes, as seen from~(\ref{b64}), whereas the constraints~(\ref{b65}) render auxiliary the traces of
all fields with spin\;$\geq$\;2;
these ensure that each field propagates the correct number of DoFs. Causal propagation of the fields is also manifest
from the dynamical equations~(\ref{b63}). Therefore, the above system consistently describes a coupled set of massive fields,
all having even-integer or odd-integer spins, propagating simultaneously in a constant EM background.

The $d$-dimensional action that would give rise to the system~(\ref{b63})--(\ref{b65}) is guaranteed to exist, and it will be
free of negative-norm states, as has been shown in Section~\ref{sec:noghost}. As in the spin-2 case, by proper redefinitions
of the fields, one expects to be able to show that the various fields are coupled only in the presence of a non-trivial background.

\vskip 36pt

%%%%%%%%%%%%%%%%%%%%%%%%%%%%%%%%%%%%%%%%%%%%%%%
\scs{Concluding Remarks}\label{sec:Conclusion}
%%%%%%%%%%%%%%%%%%%%%%%%%%%%%%%%%%%%%%%%%%%%%%%

In this paper, we have constructed a model that consistently describes a coupled system of a spin 2 and a scalar,
propagating in a constant EM background in arbitrary dimensions less than 26. Our Lagrangian propagates the correct DoFs within the
light cone, and does not contain negative-norm states. The generalization for higher spins has also been outlined.

These models do make sense in the regime of validity of a local effective action. For charged fields of mass $m$
this is when the EM field invariants are smaller than $\mathcal{O}(m^4/e^2)$, so that
instabilities~\cite{instability} are absent. Consistency in such models is achieved because of the presence of additional dynamical
fields, having the same mass as the original spin $s$ field, together with suitable non-minimal terms. The extra DoFs can be
viewed as ``new physics" on top of the system of a single spin $s$, which show up at a scale well below $m\,e^{-1/(2s-1)}$
$-$ the cutoff upper bound reported in~\cite{PR2}. All the dynamical fields present have gyromagnetic ratio equal to 2,
in accordance with the conclusion in~\cite{FPT}.

It is curious to notice that one can add a correction to the field redefinition~(\ref{redef}) that decouples
the spin-2 field from the scalar at all orders in $G$, at the level of EoM. The required field redefinition is
\bea
\mathfrak h_{mn}&\rightarrow&\mathfrak h_{mn}\,-\,\frac{\left[\,(4+\text{Tr}G^2)^2+12\,\text{Tr}G^2\,\right]
\eta_{mn}\phi-(4+\text{Tr}G^2)\,\mathcal D_{(m}\mathcal D_{n)}\phi}{(d-1)(4+\text{Tr}G^2)^2+12(d+1)\,\text{Tr}G^2}
\nonumber\\&&~~~~~~-\,\frac{3i\left[\,3(G\cdot\mathcal D)_{(m}\mathcal D_{n)}-\mathcal D_{(m}(G\cdot\mathcal D)_{n)}
\,\right]\phi}{(d-1)(4+\text{Tr}G^2)^2+12(d+1)\,\text{Tr}G^2}.
\eea{newredef}
It reduces
Eqs.~(\ref{a100})--(\ref{a104}) to \beq \left(\mathcal{D}^2-2-\tfrac{1}{2}\text{Tr}G^2\right) \mathfrak h_{mn}
-2i(G\cdot\mathfrak h-\mathfrak h\cdot G)_{mn}=0,\qquad \mathcal D^m\mathfrak h_{mn}=0,
\qquad \mathfrak h=0,\eeq{r1} \beq \left(\mathcal{D}^2-2-\tfrac{1}{2}\text{Tr}G^2\right)\phi=0.\eeq{r2}
Clearly, the two fields are completely decoupled from each other.

However, the redefinition~(\ref{newredef}) cannot decouple them at the Lagrangian level. This is clear from
the fact that~(\ref{newredef}) does not redefine the scalar $\phi$, so that the terms quadratic in the spin-2 field
(and trace thereof), obtained after the field redefinition, are identical to the first two lines of~(\ref{model}).
But the latter are simply the spin-2 Argyres-Nappi Lagrangian, which holds good in no dimensions other than 26.
The conclusion is that there must be cross-couplings between the fields in $d\neq26$, because mere field redefinitions
of the spin-2 field alone cannot affect the (in)consistency of a model.

There remains the possibility, which we leave as future work, that a
redefinition of the scalar field may decouple the fields at the Lagrangian level as well. One would then have a
consistent model for a single charged spin 2 in any dimension less than 26, which could be useful, among others, for
constructing holographic models of $d$-wave superconductors~\cite{d-wave}.

We conclude with some comments on how our results fit into the existing higher-spin gauge theory literature.
One can use a gauge invariant formulation to construct consistent EM interactions of a massive spin 2 field, as it
was done in~\cite{Zinoviev}. Moreover, the $1-s-s$ cubic vertices in~\cite{Boulanger,Taronna} are directly relevant
to the present paper, since they encode interactions of a spin-$s$ field with a $U(1)$ gauge field. One natural
question about them is whether they reproduce the cubic vertices presented here, where the $U(1)$ is treated as a
background. Consider for instance the $1-s-s$ cubic vertices for massive spin-$s$ fields written down in~\cite{Taronna}.
They were derived from string theory in 26 space-time dimensions. When the $U(1)$ field is an external source with constant
field strength, one immediately finds that, up to total derivative terms, the only vertices that survive are dipole terms
that contain no derivatives; such vertices cannot be field redefined away. This is in complete agreement with
the finding of~\cite{PRS}, that the cubic vertices in the 26-dimensional Argyres-Nappi Lagrangian~(\ref{AN}) contain only
dipole terms up to field redefinitions. A dimensional reduction, as in Eqs.~(\ref{label})--(\ref{Dsplit}), would thus produce
cubic vertices that contain no derivatives. The full interacting theory may still contain higher derivative cubic
terms, which vanish up to field redefinitions when the $U(1)$ field strength is treated as a constant background. To study unitarity
in the presence of such vertices, a derivation from string theory is probably a crucial tool, as it has been in our analysis.

\vskip 36pt

%%%%%%%%%%%%%%%%%%%%%%%%%%
\section*{Acknowledgments}
%%%%%%%%%%%%%%%%%%%%%%%%%%

We would like to thank A.~Sagnotti and M.~ Taronna for stimulating discussions and useful comments. MP is supported
in part by the NSF grant PHY-0758032 and by the ERC Advanced Investigator Grant no.\,226455 ``Supersymmetry,
Quantum Gravity and Gauge Fields" (SUPERFIELDS), while RR is supported in part by Scuola Normale
Superiore, by INFN and by the ERC Advanced Investigator Grant no.\,226455 ``Supersymmetry, Quantum Gravity
and Gauge Fields" (SUPERFIELDS).

\end{document}